\documentclass[11pt,twoside]{article}

\usepackage{baaa}

\usepackage{graphicx}
\usepackage{subfigure}
\usepackage{amssymb}
\usepackage[spanish,activeacute,english]{babel}
\usepackage[latin1]{inputenc}
\usepackage{verbatim}
\usepackage{amsmath}
\usepackage{amsfonts}
\usepackage{amssymb}
\usepackage[colorlinks=true,dvips]{hyperref}
\usepackage{natbib}

\begin{document}
\myselectenglish
\vskip 1.0cm \markboth{Bengochea G. R.}%
{Observational challenges in dark energy models}

\pagestyle{myheadings}

\vspace*{0.5cm}

\noindent PRESENTACI\'{O}N ORAL

\vskip 0.3cm
\title{Observational challenges in dark energy models}

\author{G. R. Bengochea$^{*}$}

\affil{%
  (*) Instituto de Astronom\'{\i}a y F\'{\i}sica del Espacio (CONICET-UBA)\\
}

\begin{abstract}
Cosmological distances inferred from supernova Ia observations
constitute the most direct and solid evidence for the recently
detected accelerated expansion of the universe. In this
contribution, we show some inconsistencies between two of the main
light-curve fitters used for the elaboration of supernova Ia data
sets, opening new observational challenges regarding the use of
these luminosity distances when combined with CMB and BAO data. We
also mention ongoing analysis related to alternative models. The
resolution of these challenges will be crucial for XXI century
cosmology.
\end{abstract}

\begin{resumen}
Las distancias cosmol\'{o}gicas inferidas a partir de observaciones de
supernovas del tipo Ia constituyen la evidencia m\'{a}s directa y
s\'{o}lida de la recientemente detectada expansi\'{o}n acelerada del
universo. En esta contribuci\'{o}n, se muestran algunas
inconsistencias entre las dos maneras m\'{a}s usadas para procesar las
curvas de luz de las supernovas Ia, abriendo nuevos desaf\'{\i}os
observacionales referidos al uso de estas distancias luminosas
cuando son combinadas con datos del CMB y BAO. Tambi\'{e}n se
mencionan an\'{a}lisis en curso, relacionados con modelos
alternativos. La resoluci\'{o}n de estos desaf\'{\i}os ser\'{a} crucial para la
cosmolog\'{\i}a del siglo XXI.
\end{resumen}

\section{Introduction}
\label{intro}

The discovery of the accelerated expansion of the universe in 1998
through distant supernova Ia (SN Ia) observations was awarded with
the Nobel Prize in Physics in 2011. Assuming that the universe, at
large scale, is correctly described by an FRW cosmology, these
observations of luminosity distances yielded the most conclusive
evidence for the need of adding an extra component with negative
pressure to the model, namely dark energy, responsible of the
acceleration. The combination of these observations with, for
instance, the cosmic microwave background (CMB) and the baryon
acoustic oscillations (BAO) ones, support today the
\emph{concordance model} ($\Lambda$CDM) according to which 72\% of
the energy density of the universe is dark energy (Amanullah et
al. 2010).

The flux measurements (or apparent magnitudes) in different epochs
and distinct passbands are processed with the so called
\emph{light-curve fitters} to obtain luminosity distance values.
The two most used methods are MLCS and SALT2 (e.g. Jha et al. 2007
and Guy et al. 2007). It is important to remark that a 0.2
apparent magnitude difference leads to a 10\% error in the
luminosity distance value. While MLCS only uses the nearby
supernovae to calibrate its empirical parameters, SALT2 uses the
whole data set. But, when using the complete data set, SALT2 is
forced to adopt a cosmological model for those supernovae that lie
beyond the range where the linear approximation to the Hubble law
is valid. Typically flat models are assumed ($\Omega_k=0$),
$\Lambda$CDM or $w$CDM ($w=const$, being $w$ the dark energy
equation of state. The particular case of $w=-1$ corresponds to a
cosmological constant $\Lambda$). Therefore, the distances
inferred with SALT2 hold a degree of model dependence. Several
authors have already pointed out that the obtained values for some
cosmological parameters (for example $w$) differ significantly
depending on the fitter applied (e.g. Kessler et al. 2009,
Sollerman et al. 2009 and Komatsu et al. 2011).

\section{How flat is the universe?}
In the framework of a $\Lambda$CDM model, allowing $\Omega_k$ to
vary, we studied the confidence intervals using the whole SNe Ia
data set from SDSSII (Kessler et al. 2009) processed with MLCS and
with SALT2. The analysis with MLCS (Fig.1, {\it Right}) showed
that the flat case ($\Omega_m=0.27$) lies outside the 3$\sigma$
confidence level, while with the same data set, but processed with
SALT2 this does not happen (Fig.1, {\it Left}).

\begin{figure}[h]
\centering
\includegraphics[width=0.65\textwidth]{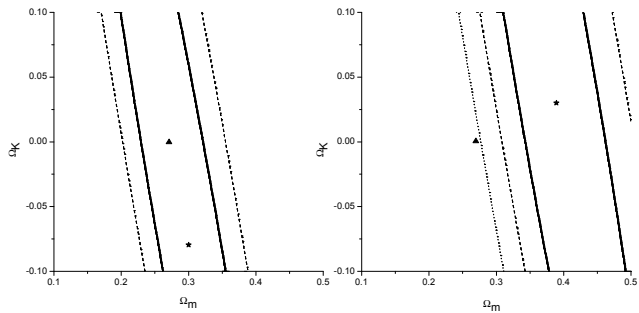}\caption{SDSSII SN Ia data set in the $\Lambda$CDM model
framework. {\it Left:} Confidence intervals at 68.3\% and 95\% for
SALT2. {\it Right:} Confidence intervals at 68.3\%, 95\% and
99.7\% for MLCS. Best fits are indicated with a star whereas the
standard flat $\Lambda$CDM ($\Omega_m=0.27$) is marked with a
triangle.} \label{flatout}
\end{figure}

Since the responsible of this fact is the fitter and not the SNe
Ia, one could wonder which SNe Ia set to use to be combined with,
for example, CMB data, which leave little margin to the variation
of $\Omega_k$ (e.g. Komatsu et al. 2009). Looking at Fig. 1, one
would choose the data set processed with SALT2; however we ought
to remember that SALT2 retains a degree of model dependence
because typically a flat $\Lambda$CDM model is assumed. Some
authors have remarked that imposing $\Omega_k=0$ could bring
serious problems when reconstructing the equation of state $w$ of
the dark energy. Omitting only a 2\% of curvature leads to the
reconstruction, employing luminosity distances data, to yield very
physically different results than when using $H(z)$ values
(Clarkson et al. 2007). This is the result of the well known
$\Omega_k-w$ degeneration (e.g. Spergel et al. 2007).

\section{Tension between data sets or fitters?}

In Wei (2010), a tension is found between SN Ia data sets, and
also between the later with BAO and CMB. The tension was
attributed to certain supernovae and the author proposed a
truncation method to remove the outliers and release the tension.
Guided by this finding, we found something similar between the
Union2 (Amanullah et al. 2010) and SDSSII SN Ia data sets when
combined with BAO/CMB data according to Sollerman et al. (2009).
But the interesting thing was to find that this tension only
appears when the data sets are processed with different fitters.
Therefore, the tension is not between SN data sets but between the
light-curve fitters used. We also found that using the same
truncation method, the same SNe Ia data set behaves in different
ways just for having been processed with one or another fitter, as
if they were two different data sets (Bengochea 2011).

\begin{figure}
  \centering
  \hfill\begin{minipage}[b]{.4\textwidth}
    \centering
    \includegraphics[width=\textwidth, origin=c, angle=0]{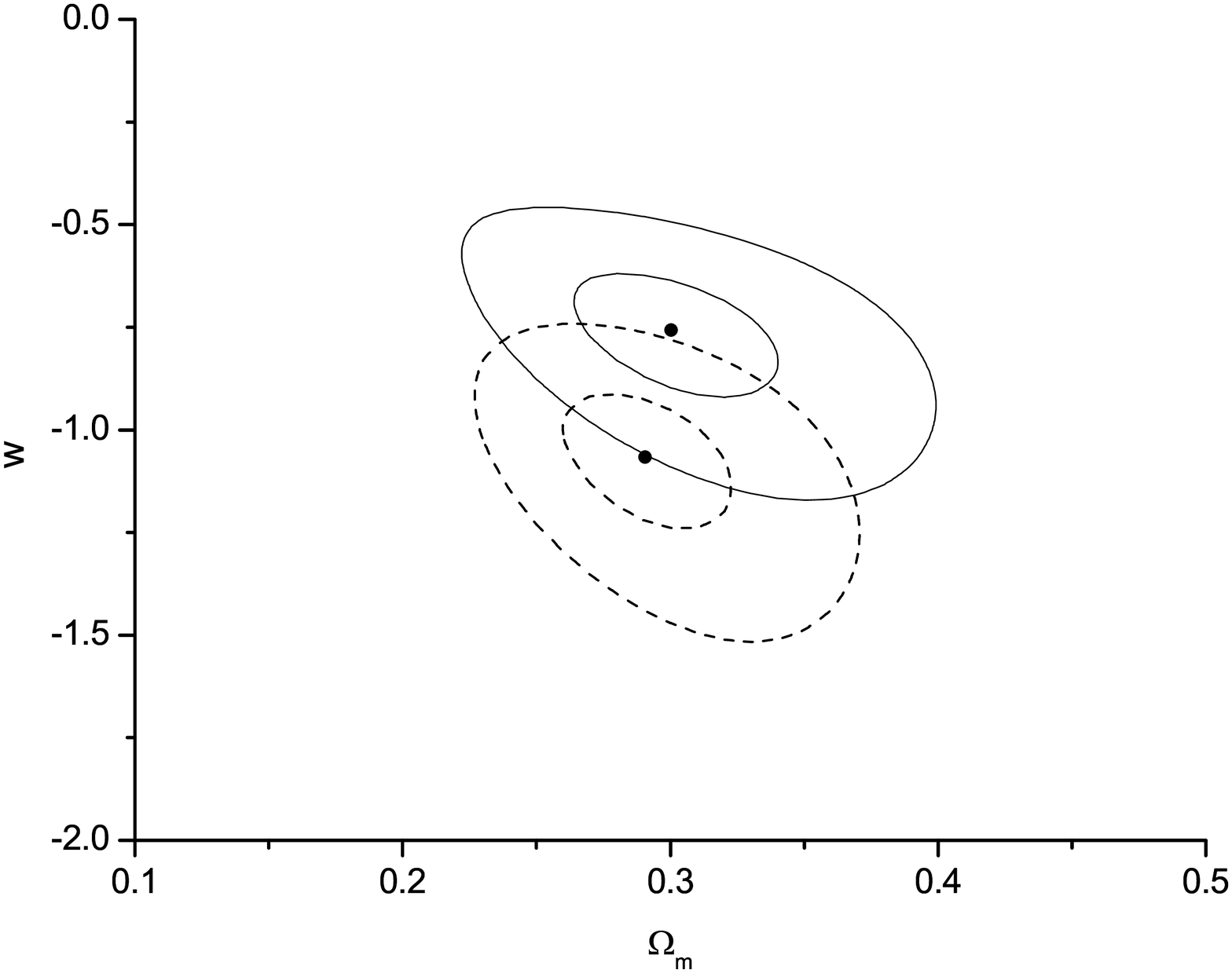}
  \end{minipage}~\hfill%
  \begin{minipage}[b]{.4\textwidth}
  \centering
    \includegraphics[width=\textwidth, angle=0]{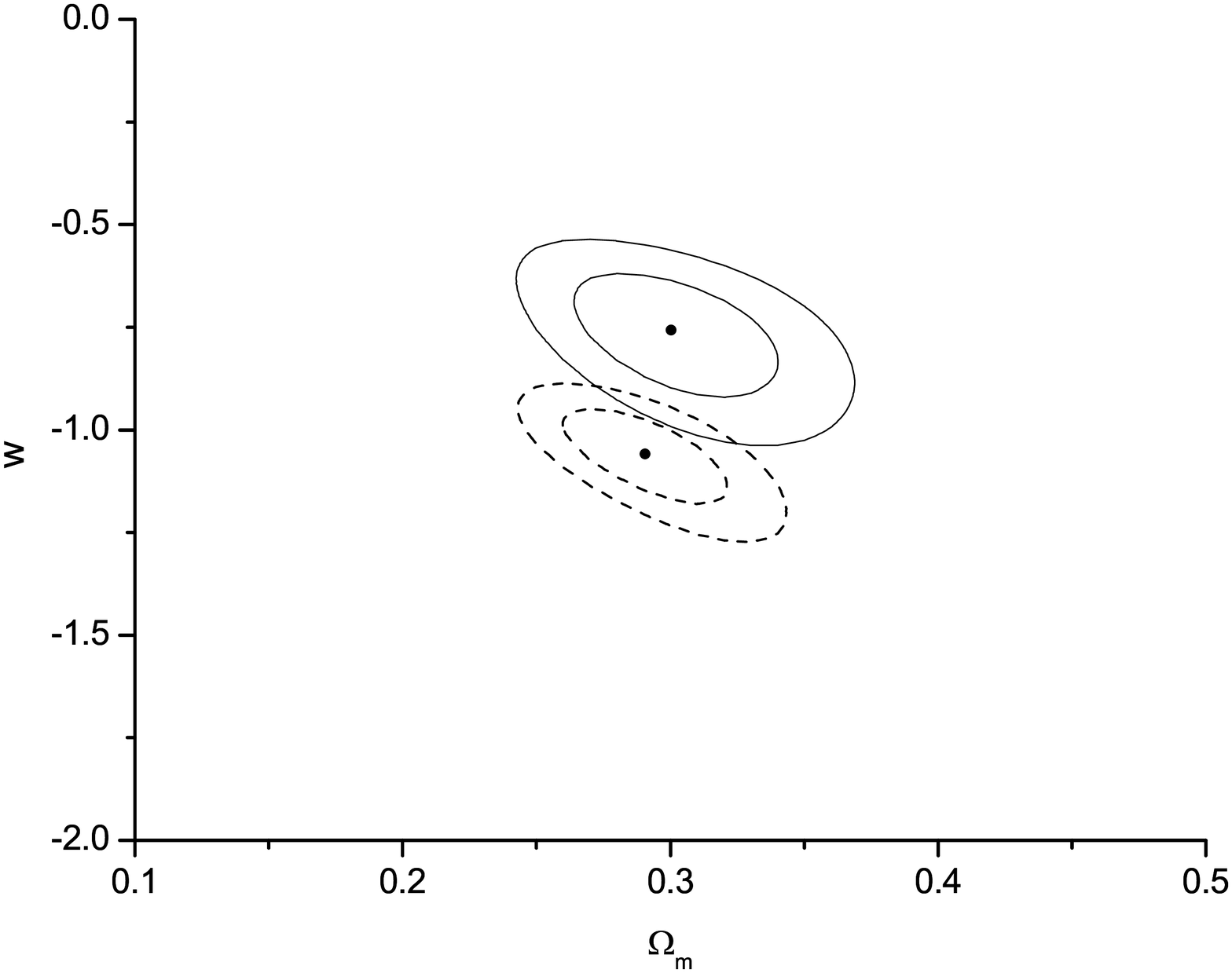}
  \end{minipage}~\hfill~\\[-10pt]
  \hfill\begin{minipage}[t]{.45\textwidth}
    \caption{Confidence intervals at 68.3\% and 99.7\% coming from combining SN Ia and BAO/CMB data.
    Solid lines correspond to SDSSII (SALT2) whereas dash lines correspond to SDSSII (MLCS).}
    \label{fig3}%
  \end{minipage}~\hfill%
  \begin{minipage}[t]{.45\textwidth}
    \caption{Confidence intervals at 68.3\% and 95\% from combining SNe Ia + BAO/CMB. Union2 (SALT2-dashed lines) vs SDSSII (MLCS-solid lines).}
    \label{fig2}%
  \end{minipage}\hfill~%
\end{figure}

The comparison of the Figures 2 and 3 allows to appreciate how two
light-curve fitters used for the same SN Ia data set produce the
same result as two distinct SN Ia data set.

One of the main goals of the future observational projects will be
to figure out if dark energy is a cosmological constant or
something more exotic ($w\neq-1$). In the search for such a
characterization, \emph{phantom} models ($w<-1$) in which the
energy density can become infinite at a finite time driving to a
\emph{big rip}, have not yet been observationally discarded
(Caldwell 2002). Bearing this in mind, another interesting result
was to find that, for several alternative theories, the fitters
add an extra degeneration in the results, favoring for example
that when combining SN Ia with BAO/CMB the equation of state $w$
is phantom or not, depending on the fitter used (Table 2 of
Bengochea 2011), or that certain parametrization for the dark
energy predict a future deceleration or not (Li et al. 2010).

Some improvements to reduce this kind of systematic errors and
others not mentioned here, have been developing (e.g. Sullivan et
al. 2010 and 2011, Marriner et al. 2011). Following what was
showed by Sollerman et al. (2009), the study of how this kind of
degeneration between light-curve fitters affects the result when
SN Ia data sets are used to put observational constraints to
inhomogeneous models is in progress (Bengochea 2012), being these
later a very interesting subject to study, as an alternative to
the FRW case, because of the capability of describing a vast
variety of observations very well \emph{without} dark energy.

\section {Conclusions}

While assuming a FRW-$\Lambda$CDM cosmology the evidence about the
existence of dark energy seems undisputed, as of today the two
most used light-curve fitters for the elaboration of the
luminosity distance SN Ia data sets present inconsistencies
between them, making the very same data set to behave, under
certain circumstances, as two different ones, generating false
tensions between SN Ia data, CMB and BAO. The extra degeneration
produced in the analysis by these fitters make more difficult to
characterize the equation of state $w$ of dark energy. It is
precise to elaborate calibration methods that are independent of
the same cosmological models being evaluated or the prior
$\Omega_k=0$, get to better understand the evolution of the
metallicity with the redshift, the variability of the intrinsic
color and the dust extinction and other factors that allow to
reduce systematic error sources. Some authors have already begun
to work on this. On the other hand, the impact of the differences
between the light-curve fitters must be also understood when SN Ia
data sets are used to put constraints to free parameters of
alternative models to the FRW case.

\acknowledgments G.R.B. is supported by CONICET.

\references

\texttt{Amanullah R. et al. 2010, ApJ, 716, 712.}

\texttt{Bengochea G. R. 2011, PLB, 696, 5.}

\texttt{Bengochea G. R. 2012, in progress.}

\texttt{Caldwell R. R. 2002, PLB, 545, 23. }

\texttt{Clarkson C. et al. 2007, JCAP, 0708, 11.}

\texttt{Jha S. et al. 2007, ApJ, 659, 122 (MLCS); Guy J. et al.
2007, A\&A, 466, 11 (SALT2).}

\texttt{Kessler R. et al. 2009, ApJS, 185, 32.}

\texttt{Komatsu E. et al. 2009, ApJS, 180, 330; 2011, ApJS, 192,
18.}

\texttt{Li Z. et al. 2010, JCAP, 1011, 31.}

\texttt{Marriner J. et al. 2011, ApJ, 740, 72.}

\texttt{Sollerman J. et al. 2009, ApJ, 703, 1374.}

\texttt{Spergel D. N. et al. 2007, ApJS, 170, 377.}

\texttt{Sullivan M. et al. 2010, MNRAS, 406, 782; 2011, ApJ, 737,
102.}

\texttt{Wei H. 2010, PLB, 687, 286.}

\end{document}